\documentclass{aastex}
\usepackage{emulateapj5}

\def \irasa{IRAS~13349+2438\/}
\def \irasb{IRAS~14026+4341\/}
\def\HST{{\it HST\/}}
\def\deg{\ifmmode^\circ\else$^\circ$\fi}

\slugcomment{Submitted to  Astrophysical Journal}

\shorttitle{Hines et al.}
\shortauthors{FOS Spectropolarimetry of Buried QSOs}

\received{2001 April 25}
\begin{document}

\title{{\it HST} ULTRAVIOLET AND GROUND-BASED OPTICAL 
SPECTROPOLARIMETRY OF IRAS QSOS: DUSTY SCATTERING IN LUMINOUS 
AGN\altaffilmark{1}}

\author{Dean C. Hines\altaffilmark{2,3}, Gary D. Schmidt\altaffilmark{2}, 
Karl D. Gordon\altaffilmark{2}, Paul S. Smith\altaffilmark{2},
Beverley J. Wills\altaffilmark{4}, Richard G. 
Allen\altaffilmark{2}, and Michael L. Sitko\altaffilmark{5}}

\altaffiltext{1}{Based on observations with the 
NASA/ESA Hubble Space Telescope obtained at the Space Telescope 
Science Institute, which is operated by the Association of 
Universities for Research in Astronomy, Inc., under NASA contract 
NAS5-26555.}

\altaffiltext{2}{Steward 
Observatory, The University of Arizona, Tucson, AZ 85721; dhines, 
gschmidt, kgordon, psmith, 
rallen@as.arizona.edu}
\altaffiltext{3}{Guest Observer, McDonald 
Observatory, The University of Texas at Austin }
\altaffiltext{4}{Astronomy Department 
\& McDonald Observatory, The University of Texas, Austin, TX 
78712; bev@panic.as.utexas.edu}
\altaffiltext{5}{Dept.  of Physics, 
University of Cincinnati, Cincinnati, OH 45221-0011; 
sitko@physics.uc.edu}

\begin{abstract}
We present UV and optical spectropolarimetry of two highly polarized 
{\it IRAS\/}-selected QSOs, \irasa\ and the BALQSO \irasb.  The 
polarization in both objects rises rapidly toward the blue, peaks near 
3000\AA\ in the rest frame and remains nearly constant for shorter 
wavelengths.  The rest frame optical polarized flux density spectra 
also increase rapidly towards the blue, but then decrease dramatically 
below 3000\AA. This distinctive wavelength dependence of polarized 
flux shows that the polarization is produced by dust scattering.  As 
for many Seyfert, radio and Hyperluminous Infrared Galaxies (HIGs), 
the lower polarization of the weak 
[\ion{O}{3}]$\lambda\lambda4959,5007$ lines in \irasa\ suggests that 
the scattering grains lie interior to, or mixed with the narrow line 
gas.  We construct full radiative transfer models of these systems 
consisting of a dusty sphere of modest optical depth illuminated 
axisymmetrically from within by a powerlaw QSO spectrum.  We show that 
this simple model successfully reproduces the qualitative polarization 
properties of the objects.  Despite similarities to other {\it 
IRAS\/}-selected BALQSOs, our FOS spectropolarimetry of \irasa\ does 
not reveal broad absorption lines.  \irasb\ has an \ion{Al}{3} BAL in 
both scattered and total flux density.  We discuss these two objects 
in terms of both orientation and evolutionary unified schemes for 
QSOs, BALQSOs and HIGs.
\end{abstract}


\section{INTRODUCTION}

Non-stellar radiation processes have been observed in the nuclei of 
galaxies for many years, but the fundamental nature and structure of 
the central engines have remained elusive.  In the past two decades, 
it has become clear that many of the apparently disparate properties 
of the active galactic nuclei (AGNs) are illusions caused by 
obscuration and orientation.  We are beginning to use these 
differences to probe the detailed 3-dimensional structure of the 
central engines.  In particular, the {\it IRAS\/} mission found a 
number of galaxies with IR luminosities and far-IR colors that are 
consistent with the most luminous AGNs (i.e., the QSOs).  Detailed 
investigations of these objects, primarily through spectro- and 
imaging polarimetry, reveal QSO-like spectra in polarized light (e.g., 
Wills et al.  1992; Hough et al.  1993; Hines \& Wills 1993, 1995; 
Goodrich et al.  1996; Hines et al.  1995; Young et al.  1996a,b; 
Wills \& Hines 1997; Tran et al.  2000).  In a few objects, the 
presence of highly polarized ($p\ge10\%$) extended emission secures 
scattering as the polarizing mechanism (Miller \& Goodrich 1990; Tran 
1995; Hines et al.  1999; Tran, Cohen \& Villar-Martin 2000).  For the 
few objects where dichroic absorption by aligned dust grains cannot be 
ruled out entirely, the grains and aligning magnetic fields would have 
to be very different than those responsible for the interstellar 
polarization within our Galaxy.

The nature and intrinsic luminosity of the central engine of an 
obscured AGN can sometimes be inferred from the shape and strength of 
the flux spectrum in polarized light (e.g., Wills et al.  1992; Hines 
et al.  1995, 1999).  However, nearly all of these investigations have 
been limited to spectropolarimetry in the optical wavelength range.  
Since the scattering crossection of particle can be approximately 
wavelength-independent over small wavelength intervals, it can be 
difficult to distinguish between electron- and dust-scattering as the 
underlying mechanism based solely upon optical observations.  In this 
paper we present spectropolarimetry of two {\it IRAS\/}-selected QSOs 
that extends into the UV, where differences in scattering mechanisms 
are more noticeable.  The results allow us to identify dust scattering 
as the dominant polarizing mechanism in both objects, and to 
investigate through modeling the geometry of the dust surrounding the 
central engines.

\section{OBSERVATIONS}

Spectropolarimetric observations of \irasa\ and \irasb\ were conducted 
with the Faint Object Spectrograph (FOS) aboard the {\it Hubble Space 
Telescope} ({\it HST}) on UT 1995 Sep.  11 and 1996 May 8, 
respectively, using four positions of Waveplate B, the G270H grating, 
the blue channel in ACCUM mode, and the 1\arcsec\ aperture (GO 5928: 
PI Hines).  A G190H observation without the polarimetry optics was 
added for \irasa\ on UT 1995 Sep.  11 to extend the spectrum through 
\ion{C}{4} $\lambda1549$.  A similar G190H observation of IRAS 
P14026+4341 by Turnshek et al.  (1996) was retrieved from from the 
Space Telescope Science Institute archive, and is included herein.

The G190H observations were reduced using the standard STScI pipeline 
(Keyes et al.  1995).  For the spectropolarimetry data, the 
polarization state of the light reaching the FOS is altered by the two 
COSTAR mirrors used to correct for the spherical aberration in the 
primary (Storrs et al.  1998).  Light from the telescope is deflected 
by the first mirror upward and across the centerline of the telescope 
to a second mirror mounted directly ahead of the FOS aperture.  A 
spurious linear polarization signal is introduced because the mirrors 
have different reflectivities for light polarized parallel and 
perpendicular to the plane of incidence.  The parallel and 
perpendicular polarizations also differ in phase, resulting in some 
conversion of linear into circular polarization and vice versa.  
Fortunately, the effects of the mirrors have been characterized by 
observing both polarized and unpolarized standard stars, and the 
polarization of an unknown source can be recovered by inverting the 
reflection process (Storrs et al.  1998).  Since only four waveplate 
positions were used for the present observations, we cannot determine 
the circular component for these two QSOs, but it is not expected to 
be significant.  The effect of linear to circular conversion would 
represent a small error in the measured percentage polarization, 
$p_{\rm obs}\ge0.929\times p_{\rm real}$, and the close match to the 
ground-based data near 3200\AA\ implies that this correction must be 
very small.

The \HST\ data are complemented by ground-based spectropolarimetry 
obtained with the CCD polarimeter on the 2.3~m Bok telescope of 
Steward Observatory and with the LCS-Spectropolarimeter attached to 
the 2.7~m Smith telescope at McDonald Observatory.  The design and use 
of the Steward\footnote{Upgraded with a thinned, AR-coated, and 
UV-enhanced $1200\times800$ Loral CCD} and McDonald instruments are 
discussed in Schmidt, Stockman, \& Smith (1992) and Goodrich (1991), 
respectively.  The analysis of data from both instruments parallels 
the description of Miller, Robinson, \& Goodrich (1988).  Polarimetric 
calibrations were made each run using a fully-polarizing element 
(either a polarizing sheet or prism) inserted into the beam, and 
position angles are registered with the equatorial system to better 
than 1$^\circ$ through observations of interstellar-polarized standard 
stars from the grid calibrated by Schmidt, Elston \& Lupie (1992).  
Total spectral flux is also available for each observation, using 
nightly calibration observations of flux standards from the {\it 
IRAF\/} database obtained with the identical instrumental setups.

The data were obtained primarily with 2-3\arcsec-wide E-W slits, and 
the spectra were optimally extracted using an 8\arcsec\ 
profile-weighted aperture.  The resulting rest-frame velocity 
resolution is $\sim$600~km~s$^{-1}$.  The McDonald spectropolarimeter 
has response to the near-UV atmospheric cut-off, but suffers from 
substantial (variable) fringing for wavelengths longer than 7200\AA. 
Because the instrument is a retrofit to an existing spectrograph, the 
beamsplitter design introduces a non-uniform illumination for 
wavelengths shorter than $\sim 4000$\AA\ (Goodrich 1991) which 
manifests itself as a small instrumental polarization ($p_{inst} \le 
3\%$ at 3200\AA) that can be removed successfully through observations 
of unpolarized standard stars (e.g., Hines \& Wills 1995).  The 
excellent match between the {\it HST} and McDonald polarimetry data 
herein shows efficacy of this correction.

The Steward spectropolarimeter is very efficient throughout most of 
the optical spectral region, but glass spectrograph elements strongly 
limit its performance below 4000\AA, just redward of the \ion{Mg}{2} 
$\lambda$2800 emission line in the rest frame of our objects.  The use 
of a Wallaston prism for the beam splitter avoids the illumination 
problems encountered with the McDonald instrument.

The optical polarization and position angles observed for \irasa\ and 
\irasb\ do not vary among the epochs, so the spectropolarimetric 
results were coadded over the numerous observational sequences.  In 
addition, the FOS flux density, percent polarization $p$ and position 
angle $\theta$ spectra match the McDonald data to within the 
uncertainties in the regions of wavelength overlap, so the these data 
were also averaged in the overlapping regions.  Finally, the run of 
$p$ and $\theta$ in the McDonald and Steward data is identical to 
within the uncertainties.  The flux density spectrum of \irasa\ is 
$\sim$10\% lower in the Steward data, but this difference is most 
likely caused by the slightly different slit widths used.  The Steward 
flux density spectrum was therefore scaled to match the McDonald data 
in the region of overlap.  We caution that a real change in flux 
cannot be ruled out.  The overall consistency among the three sets of 
observations, and with earlier broadband data (Wills et al.  1992; 
Hines 1994; Schmidt \& Hines 1999) indicates that the polarization 
properties of each object have not varied significantly over at least 
a $6-10$ year baseline.

Both \irasa\ \& \irasb\ are located in regions of the sky that show 
very low extinction from material within our Galaxy [E(B-V) $\le$ 
0.012: Burstein \& Heiles (1984); Schlegel, Finkbeiner \& Davis 
(1998)].  Therefore we have not attempted to correct for Galactic 
extinction in the following analysis, because the exact form of the 
extinction law depends on the line of sight direction, and introduces 
its own uncertainty.  Omission of a correction introduces at most a 
10$\%$ error in the absolute total and polarized flux density at 
2200\AA\ (observed).

Table 1 presents a log of the observations.  All analyses discussed 
below were performed on the complete data sets.

\section{THE UV/OPTICAL POLARIZATION SPECTRA OF \irasa\ \& \irasb}

Figures 1 \& 2 present the rest-frame spectropolarimetry results for 
\irasa\ \& \irasb.  Shown from top to bottom in each figure are: the 
electric-vector position angle $\theta_{\lambda}$; the rotated Stokes 
parameter $q^\prime_\lambda$; the total flux density $F_\lambda$; and 
the Stokes flux density spectrum\footnote{The rotated Stokes parameter 
$q^\prime_{\lambda} = q_{\lambda} \cos{2\theta}+u_{\lambda} 
\sin{2\theta}$ represents the total percentage polarization measured 
in a coordinate system aligned with the overall symmetry axis of the 
polarizing mechanism.  In practice, $\theta$ is the average measured 
position angle ($<\theta_{\lambda}>$).  For objects that display a 
small range in $\theta$ (like those discussed here), the Stokes flux 
$q^\prime_\lambda F_\lambda$ is substantially the same as the 
polarized flux $p_\lambda F_\lambda$, but avoids the bias and peculiar 
error distribution associated with $p$.}, $q^\prime_\lambda 
F_\lambda$.  For brevity we shall refer to the latter as the polarized 
flux.  The rotated Stokes parameters for \irasa\ \& \irasb\ were 
constructed using the position angle averaged over the entire 
UV/optical wavelength range to form the rotated Stokes paramter.

\begin{figure*}
\figcaption{Spectropolarimetry of \irasa.  From top to 
bottom: the position angle of polarization $\theta_\lambda$; the 
rotated Stokes parameter $q^\prime_\lambda$ for a coordinate system 
aligned with the overall polarization position angle; the total flux 
density spectrum $F_\lambda$; and the Stokes (polarized) flux density 
$q^\prime F_\lambda$.  The data have been shifted into the rest frame 
and flux densities have been multiplied by $1+z$.  No cosmological 
corrections have been applied.}
\end{figure*}

\begin{figure*}
\figcaption{Spectropolarimetry of \irasb, presented as in Figure 1.}
\end{figure*}

Both objects exhibit high polarization increasing dramatically toward 
the blue to a maximum of $p_{\rm max}\sim12\%$ near 3000\AA\ and then 
leveling off into the UV. In each case, no more than $10-15$\% of the 
light at 7000\AA\ originates from the host galaxy and the contribution 
decreases to shorter wavelengths (Wills et al.  1992; Hutchings \& 
Neff 1992; Hutchings \& Morris 1995), so the strong polarization 
wavelength dependence must be due to either the polarizing mechanism 
itself or to dilution by unpolarized light from the direct (possibly 
reddened) QSO. The bottom panels of each figure reveal perhaps the 
most noteworthy result, that in both examples, {\it the polarized flux 
spectra show a peak and strong decline for rest wavelengths 
$\lambda\le3000$}\AA. The polarized flux density spectra are 
independent of any diluting flux from unpolarized sources.  As we will 
argue in the following sections, this shape cannot be reproduced by 
electron scattering even with dust extinction, nor can it be produced 
by dichroic absorption by (magnetically) aligned dust grains like 
those in the Interstellar medium (ISM) of our Galaxy.  Instead, the 
polarization of these AGN must be produced by dust scattering.  Table 
2 summarizes the polarization properties of the two objects.  The 
polarization properties of the emission lines were derived (in Stokes 
parameters) after subtraction of the local continuum.  The data are 
not corrected for the bias inherent in polarization (see, e.g., 
Simmons \& Stewart 1985; Clarke \& Stewart 1986).

\begin{figure*}
\figcaption{Comparison between the UV total flux 
density spectra of \irasa\ and \irasb.}
\end{figure*}

Figure 3 compares the UV total flux spectra of the two sources.  Broad 
absorption lines (BALs) associated with both high-ionization 
(\ion{Si}{4}, \ion{C}{4}) and low-ionization (\ion{Mg}{2}, 
\ion{Al}{3}) species are apparent in the spectrum of \irasb, 
confirming that it is a member of the class of low-ionization BALQSOs 
(e.g., Weymann et al.  1991).  Although \irasa\ has a similarly red UV 
spectrum, it shows no sign of atomic absorption in the UV/optical (but 
see next section).  The emission and absorption line properties are 
presented in Table 3, and the objects are described individually 
below.

\subsection{\irasa}

\irasa\ ($z=0.1076$) was the first previously-unidentified QSO 
discovered by {\it IRAS\/} (Beichman et al.  1986).  Wills et al.  
(1992) found that the {\it UBVRI\/} polarization is both strong and 
highly wavelength dependent, even after correction for dilution by 
unpolarized host galaxy starlight ($\le 10\%$ of light near 
H$\alpha$).  They also found that H$\alpha$ is polarized like the 
surrounding continuum, and that the position angle is nearly constant 
at $\theta\approx124\deg$ from 0.36$\mu$m to 0.8$\mu$m and a possible 
rotation to 137\deg\ at 2.2$\mu$m (Sitko \& Zhu 1991).  These 
characteristics are borne out in Figure~1.  Even though there is a 
slight position angle rotation between the broad emission lines 
($\theta\approx121\deg$) and continuum ($\theta\approx124\deg$), the 
lack of structure in $p$(\%) near these lines, and the lack of 
polarization variability imply that synchrotron emission is not the 
dominant polarizing mechanism.  Instead, the polarizing material must 
be witness to a slightly different geometry for the broad 
emission-line region (BELR) than for the continuum source.

In \irasa\ the mean polarization position angle is parallel to the 
apparent major axis of the host galaxy (Hutchings \& McClure 1990; 
Wills et al.  1992), suggesting that the polarizing mechanism 
``knows'' about the symmetry axis of the galaxy.  However, thus far no 
obvious extended polarized structures have been imaged.  In addition, 
$p(\%)$ decreases at [\ion{O}{3}] $\lambda5007$.  Reduced polarization 
in the forbidden lines is common for Seyfert galaxies (e.g., Miller \& 
Goodrich 1990; Tran 1995), for radio galaxies (e.g., di Serego 
Aligheri, Cimatti \& Fosbury 1994; Smith et al.  1997; Cohen et al.  
1999; but see di Serego Aligheri et al.  1997), and in the most 
luminous infrared galaxies (e.g., Hines et al.  1995, 1999; Tran et 
al.  2000; Smith et al.  2000).  In the typical case of polarization 
by small-particle scattering, that fact suggests the scattering 
material generally lies interior to, or mixed with, the narrow 
emission line region.  This general conclusion is supported by 
observations of a few resolved scattering and [\ion{O}{3}]-emitting 
regions (e.g. NGC 1068: Miller, Goodrich \& Mathews 1991; Capetti, 
Macchetto \& Lattanzi 1997, and references therein), but other 
geometries might produce similar effects.  High resolution imaging 
polarimetry as possibly afforded by instruments such as the Advanced 
Camera for Surveys to be installed aboard {\it HST} would further 
constrain the geometry in \irasa.

As noted by Beichman et al.  (1986) and Wills et al.  (1992), the 
optical total flux density spectrum of \irasa\ is red compared with 
typical optically-selected QSOs (e.g., the Palomar-Green, or PG, list; 
Schmidt \& Green 1983).  The FOS flux spectrum reveals an inflection 
at $\sim$3000\AA, below which it drops dramatically toward shorter 
wavelengths.  Broad emission lines from \ion{Mg}{2} $\lambda$2800, 
\ion{C}{3}] $\lambda$1909 and \ion{C}{4} $\lambda$1549 are observed 
with equivalent widths and velocity widths in the range typical of 
optically-selected QSOs, and there is no evidence of associated or 
broad absorption lines.  An {\it IUE\/} spectrum\footnote{The FOS 
continuum flux density is 30-40\% lower than the {\it IUE\/} spectrum.  
While the S/N ratio of the {\it IUE\/} data is low and some 
variability cannot be ruled out, the effect is most likely caused by 
the larger aperture used for the IUE data.  The {\it IUE\/} spectrum 
presented in Lanzetta et al.  (1993) was created from data originally 
obtained with {\it IUE} by B. Wills.} (Lanzetta, Turnshek, \& Sandoval 
1993) shows Ly$\alpha$ emission, again with no sign of BALs, but the 
signal to noise ratio is low.

The polarized flux spectrum of \irasa\ can be fitted by a power law 
from $\sim$$3000-7000$\AA\ with $\alpha_{\nu}\sim0.1$ 
($F_{\nu}\propto\nu^{\alpha_{\nu}}$).  Below the short-wavelength 
peak, the turnover is rapid ($\alpha_{\nu}\sim -5$).  If we assume 
that the intrinsic spectrum of the QSO is similar to a typical PG QSO 
and that the remarkable wavelength dependence in polarized flux is a 
result of reprocessing this intrinsic spectrum, the shape of the 
polarized flux spectrum informs us about the reprocessing mechanism.

A simple electron-scattering model using illumination by a typical QSO 
spectrum can be ruled out because the cross section for electrons is 
wavelength-independent, so the dilution-corrected percentage 
polarization would be constant and the polarized flux spectrum should 
duplicate the incident spectrum.  The $\sim$5$\deg$ rotation in 
$\theta$ between the lines and continuum might suggest two polarized 
components with slightly different position angles and wavelength 
dependencies, one which dominates the lines and the other the 
continuum.  However, we have not been able to construct such a model 
using electron scattering that simultaneously satisfies all of the 
observed polarization properties.

Dichroic absorption by aligned dust grains like those in the ISM of 
our Galaxy can impart wavelength dependence to $p(\%)$ (the so-called 
Serkowski law), however the high polarization of \irasa\ would imply a 
large extinction, $E(B-V)\ge p(\%)/9 \gtrsim 1.3$~mag, or $A_{\it 
V}\gtrsim4.3$ for $R=3.1$ (e.g., Wilking et al.  1982).  Both the 
continuum shape and Balmer decrement indicate $E(B-V)\lesssim0.3$ mag 
(Wills et al.  1992).  Moreover, the peak wavelength of interstellar 
polarization has not been observed to be shorter than $\sim$4000\AA\ 
(Martin, Clayton \& Wolff 1999).  An Interstellar Polarization (ISP) 
function is able to reproduce the overall shape of the polarized flux 
density, but this is always accompanied by a strong turnover in the 
percentage polarization as well, which is not observed.  A composite 
model with an electron scattered, wavelength-independent polarization 
component helps to soften the turnover in percentage polarization in 
the UV, but fails to reproduce the steep turnover in polarized flux in 
the UV.

On the other hand, spectral turnovers such as that depicted in Figure 
1 have been shown to arise naturally through scattering by small dust 
grains (e.g., Zubko \& Laor 2000, and references therein).  While we 
cannot rule out ISP completely, we offer small particle scattering as 
the only viable mechanism for the polarization of \irasa.  We explore 
this in detail in \S6.

\subsection{\irasb}

\irasb\ was identified by Low et al.  (1988) in a sample of objects 
selected to have the ``warm'' {\it IRAS\/} colors [0.25 $\le$ 
F$_\nu(25\mu$m)/F$_\nu(60\mu$m) $<$3] evidenced by \irasa\ and the PG 
QSOs.  Followup spectroscopy revealed an emission-line redshift 
$z=0.323$ and a possible BAL from \ion{Mg}{2} $\lambda$2800 (Low et 
al.  1989).  Subsequent \HST/FOS spectroscopy added a BAL feature 
probably associated with \ion{C}{4} $\lambda$1549 (Turnshek et al.  
1996).  The object shows high, strongly wavelength-dependent broadband 
optical polarization.  The position angle is constant at $\theta 
\approx 34\deg$ in {\it UBVR\/} bands, but may rotate to $\sim 45\deg$ 
at $I$ (Hines 1994; Wills \& Hines 1997).

Figure 2 confirms the strong wavelength dependence of polarization in 
the optical, but a flattening is present in the near-UV. Because the 
host galaxy contributes at most $\sim$10\% of the light in the red 
(Hutchings \& Neff 1992), the strong rise in $p(\%)$ must be intrinsic 
to the nuclear region of the QSO. The polarization of the broad 
emission lines is very similar to the nearby continuum (Table 3).  The 
position angle is independent of wavelength ($\theta\approx32\deg$), 
and appears unrelated to any obvious structure seen in the \HST\ 
images (Hutchings \& Morris 1995).

The polarized flux spectrum is extraordinarily steep in the optical 
($\alpha_{\nu}\approx1.2$).  For polarization by electron scattering, 
it would imply the bluest intrinsic spectrum ever observed for an AGN. 
The Balmer lines and \ion{Fe}{2} emission are clearly present in 
polarized light, but the Balmer decrement (H$\alpha$/H$\beta\sim1.9$) 
is flatter than for Case B recombination, and much smaller than is 
typical of the BLR in PG QSOs (H$\alpha$/H$\beta\sim3.1$; Thompson 
1992).  We interpret this as a manifestation of the scattering cross 
section of the material producing the polarization.  The S/N in 
polarized flux is not sufficient to test for the presence of 
\ion{Mg}{2} $\lambda$2800 emission or absorption, but \ion{Al}{3} 
$\lambda$1858 absorption is evident in both the polarized and total 
light.  It appears that the polarized and unpolarized light each pass 
through similar columns of low-ionization absorbing material.

Similar to \irasa, the polarized flux peaks at $\sim$3000\AA\ and 
decreases rapidly toward shorter wavelengths.  The arguments in favor 
of dust scattering made above for \irasa\ apply also to \irasb, but 
the more extreme wavelength dependence of \irasb\ makes an even 
stronger case for dust scattering.

\section{COMPARISON WITH OTHER LUMINOUS INFRARED AGN}

Though the wavelength dependence of polarized flux in \irasa\ and\\ 
\irasb\ is distinctive, it is not unprecedented.  Among the low-$z$, 
IR-luminous AGNs that were observed while \HST\ possessed UV 
spectropolarimetric capability are the Type 1 (broad emission line), 
highly-polarized sources Mrk~231 (Smith et al.  1995) and Mrk~486 
(Smith et al.  1997).  In Figure 4 (a-e) we compare the polarized flux 
distributions of these objects with \irasa\ and \irasb, and with IRAS 
07598+6508.  As discussed by Hines \& Wills (1995), the spectrum of 
the latter has been extended into the UV under the assumption that the 
polarized flux cannot exceed the total flux density spectrum, and the 
two curves shown denote whether or not the BALs are present in 
polarized light.

\begin{figure*}
\figcaption{Rest-frame polarized flux spectra for five 
IR-luminous QSOs: (a) IRAS 07598+6508 with the scattering model of 
Hines \& Wills (1995) extending into the UV; (b) Mrk~486 (from Smith 
et al.  1997) (c) \irasb; (d) \irasa; and (e) Mrk~231 (from Smith et 
al.  1995).  Panel (a) also depicts a composite QSO spectrum.  A power 
law fit to the continuum of the composite 
($F_{\nu}\propto\nu^{-0.75}$) is also shown in each panel, normalized 
to the polarized continuum at 6200\AA. Note the general similarity in 
shape of all the polarized spectra, but the variation in wavelength of 
the peak.}
\end{figure*}

Figure 4(a) also displays a composite QSO spectrum assembled from 
\HST\ FOS and ground-based data by Wills et al.  (1993).  We note that 
this spectrum is nearly identical to the mean spectrum compiled from 
the Large Bright Quasar Survey (LBQS; Francis et al.  1992).  A 
power-law fit to the continuum of the composite yields 
$F_{\nu}\propto\nu^{-0.75}$, and is shown in each panel scaled to 
match the polarized flux density spectra of each object at 6200\AA. 
The optical polarized flux densities of all five objects are 
consistent with or slightly bluer than the composite QSO total flux 
spectrum.  None of the objects is fit by the power law in the UV, as 
all show a turnover in the range $\sim$$2500-4000$\AA\ and a dramatic 
decline to shorter wavelengths.

While the peak wavelength of polarized flux density varies among the 
objects, the similarity in overall shape is striking.  The constancy 
of $\theta$ across each spectrum (not shown), with at most a 10$\deg$ 
rotation between broad emission lines and the continua, argues that 
each is dominated by a single polarizing mechanism, with possibly 
slightly different geometries applying to light from the compact 
continuum source {\it vs.\/} the BELR. Strong arguments have been 
presented in the literature for each case for polarization by dust 
scattering, and it would appear that the turnover in polarized flux in 
the near-UV is a signature of this mechanism.

Of course, counterexamples also exist.  Observations of the 
prototypical ``polarized'' Seyfert 2 galaxy NGC 1068 show that $p(\%)$ 
rises toward shorter wavelengths, and, like \irasa\ and \irasb, levels 
off around 3000\AA\ rest.  However, the object does not exhibit the 
strong turnover and decline in polarized flux to the UV (Code et al.  
1993; Antonucci, Hurt \& Miller 1994) despite the fact that spatially 
resolved optical spectropolarimetry shows that scattering by dust and 
electrons dominates the NLR and close-in nucleus respectively (Miller, 
Goodrich \& Matthews 1991).

Ultraviolet spectropolarimetry is also available for higher-redshift 
objects, including radio galaxies (e.g., Cimatti et al.  1996; 
Kishimoto et al.  2001, and references therein) and the most powerful 
Hyperluminous Infrared Galaxies (HIGs: Hines et al.  1995, 1999; 
Goodrich et al.  1996).  Again, these fail to show a turnover in 
polarized flux in the UV. In the latter cases, the data extend only to 
$\sim$2500\AA\ (rest), but the detailed agreement between a standard 
QSO composite and the polarized flux spectra rules out dramatic 
turnovers like those observed for \irasa\ and \irasb.  The extent of 
the polarized emission of the HIGs on the sky (several kpc) favors 
dust scattering, since the implied number density of scattering 
electrons would yield (unpolarized) recombination line emission in 
excess of that observed.

A primary difference between the IR-luminous AGN of Figure 4 on the 
one hand, and NGC 1068 and the HIGs on the other, is the dominance by 
host galaxy starlight and narrow emission lines in the spectra of the 
latter objects.  Whereas these (Type 2) objects reveal a QSO-like 
(Type 1) spectrum only in polarized light, all of the AGN with UV 
turnovers in polarized flux so far show a Type-1 spectrum in total 
flux.  Within the context of the orientation-dependent unified scheme, 
Type 2 sources are generally thought to be viewed at high 
inclinations, so any central (compact) scattering regions would be 
largely hidden by the obscuring torus and the polarized light would be 
dominated by near right-angle scattering off clouds far from the 
nucleus.  This is consistent with the large extent of polarized 
regions observed around the HIGs.  A higher optical depth for the 
larger scattering angles might result in reduced wavelength 
dependence, mimicking more closely the scattered spectrum from 
electrons (Kishimoto et al.  2001).  Thus, the UV turnover in 
polarized flux might not only serve as an indicator of dust 
scattering, but possibly also aid in diagnosing the scattering 
geometry.  We explore these possibilities in the following section.

\section{DUST SCATTERING IN \irasa\ \& \irasb}

We have used the DIRTY radiative transfer code (Gordon et al.  2001) 
to test if dust scattering can reproduce the qualitative behavior of 
the polarized flux density spectra shown in Figure 4, and in 
particular the observed spectra of \irasa\ and \irasb.  The DIRTY 
model computes the radiative transfer of Stokes parameterized photons 
through arbitrary distributions of dust using Monte Carlo techniques.  
Full details of this method are given by Gordon et al.  (2001) and 
Misselt et al.  (2001).

For the present application, we restricted our attention to a geometry 
similar to that invoked for unified schemes of Seyfert galaxies (e.g., 
Antonucci 1993), which has also been shown to apply to some HIGs 
(Hines et al.  1995; 1999).  Central illumination was provided by a 
point source, but photons were only allowed to ``escape'' 
axisymmetrically into a bicone aligned along the $z$-axis of the 
system.  Both cones contribute to the scattered light.  Cones with 
half-opening angles of $\theta_{\rm c} =30\degr$, $45\degr$, and 
$60\degr$ were chosen to simulate either an isotropically emitting 
source surrounded by an optically and geometrically thick torus, or a 
central source which is inherently axisymmetric.  The scattering dust 
was distributed homogeneously in a surrounding sphere, and 
calculations performed with a range of radial $V$-band optical depth 
$\tau_V=0.25-4.0$.  Dust grains were assumed to be described by either 
average Milky Way (MW) dust with $R_V=3.1$ (Cardelli, Clayton \& 
Mathis 1989; Clayton et al.  2000) or Small Magellanic Cloud (SMC) bar 
dust (Gordon \& Clayton 1998; Clayton et al.  2000) and are spherical.  
Each computational run consisted of a series of models with 
inclinations ranging from $i=0\degr$ to $90\degr$ in $10\degr$ 
increments, with the properties of an unresolved source constructed by 
integration along the line of sight.

\begin{figure*}
\epsscale{2.2}
\plottwo{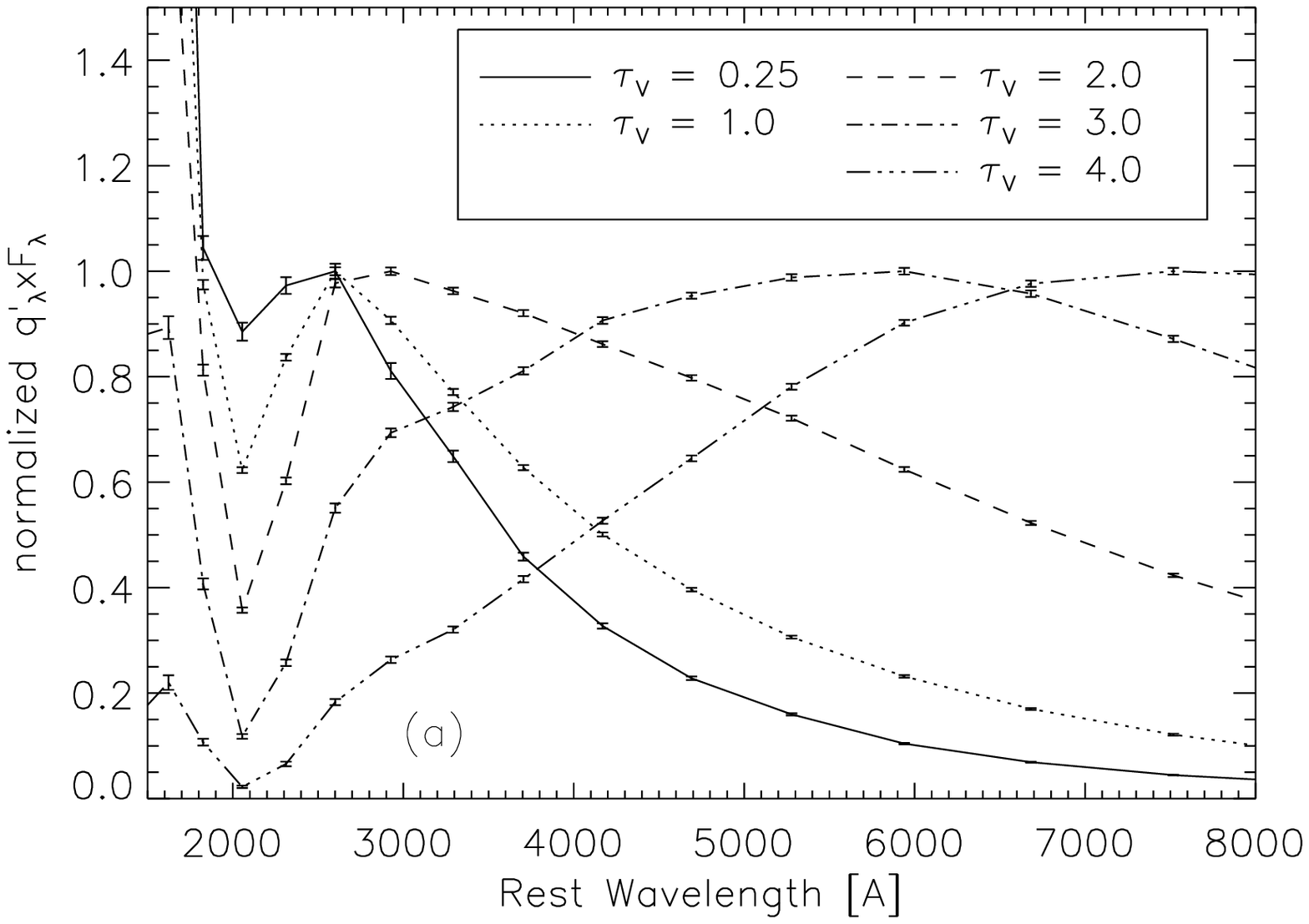}{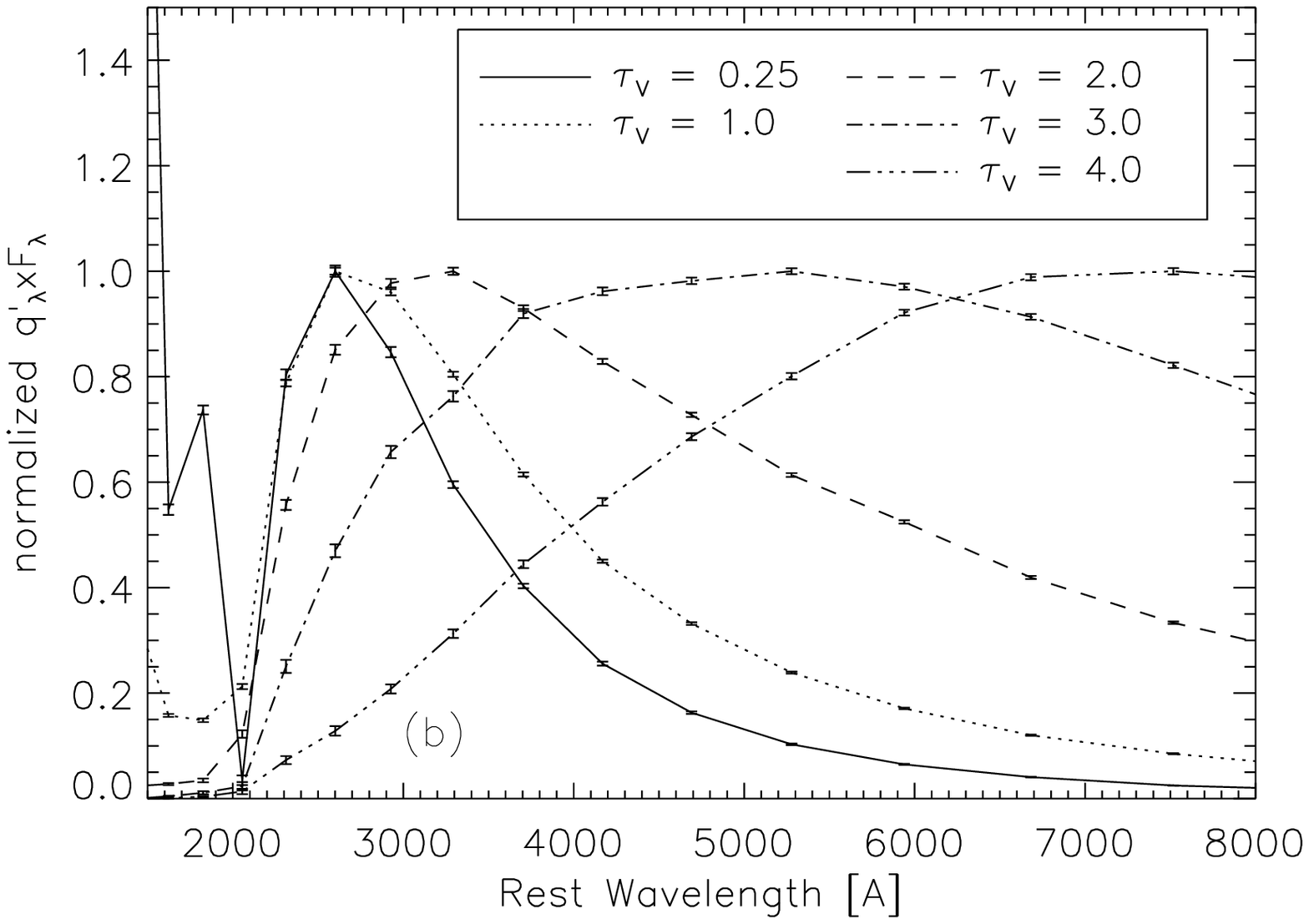}
\figcaption{Polarized flux spectra plotted for DIRTY 
models with three different radial optical depths $\tau_V$ and Milky 
Way-type dust (a) or SMC-like dust (b).  All five models assume an 
opening angle of 45$\degr$, inclination $i=40\degr$, and an intrinsic 
spectral energy distribution with $F_\nu\propto\nu^{-1}$.}
\end{figure*}

Sample results are shown in Figures 5(a \& b) as polarized flux 
density spectra for various amounts of MW and SMC-type dust, 
respectively.  The models were constructed assuming parameters similar 
to those inferred for \irasa\ by Wills et al.  (1992).  Namely: 
$\theta_{\rm c}=45\deg$ and $i=40\deg$.  Models with $i>\theta_{\rm 
c}$ are excluded in this initial model since we would not see the BELR 
directly.  In addition, models with $i\lesssim40\deg$ were unable to 
produce the high net polarizations, $p\ge15\%$, that are observed in 
these objects.

Full scattering matrices have not been tested very well, especially in 
the UV for the types of dust we are modeling.  We used dust grain size 
distributions which have been fit to the extinction curves for the MW 
and SMC. These fits use ``astronomical'' indices of refraction.  By 
``astronomical'' we mean that they have been modified from the 
laboratory measured indices of refraction to fit the observed 
extinction curve.  The actual scattering parameters (albedo and 
scattering phase function) have not been fit directly, but are derived 
from the fit to the extinction curve (a sum of the absorption and 
scattering).  The extra minima in Figure 5(b) are possibly due to the 
size distribution of dust grains (which is not forced to be smooth - 
it has wiggles).  The large polarized flux at the shortest wavelengths 
in Figure 5(a) is again a size distribution effect.  The difference 
between the SMC and MW extinction curves is entirely due to different 
dust grain size distributions.  We emphasize that this is probably an 
artifact of the way in which the extinction curves are modeled, not 
necessarily what is really happening.  The plots of Manzini et al.  
(1996) for different sized dust grains are a good visual clue to 
understanding the shape of the polarized flux spectrum.  We plan to 
explore many geometries and grain models for more general cases in a 
future paper.

It is clear that, for $\tau_V\sim1-2$, a UV turnover in polarized flux 
is indeed produced.  The turnover is basically an optical depth 
effect, with the peak representing the wavelength of maximum 
single-scattering efficiency.  For MW dust, this is influenced 
strongly by the presence of the $\lambda$2175 graphite extinction 
feature.  Thus, both the peak wavelength and the width of the peak 
increase with $\tau_V$.  In this context, we draw attention to the 
data for \irasb\ {\it vs.\/} Mrk~231 in Figure 4.  The qualitative 
behavior of the scattering models is independent of the choice of 
dust, and the shortest wavelength for the peak is found to be 
$\sim$2500\AA.

In focusing particularly on \irasa\ and \irasb, we searched for models 
that were consistent with the observed wavelength dependence of 
polarized flux, produced an (undiluted) degree of polarization at 
least as large as observed, and a scattered plus unscattered flux 
density no brighter than the observed total flux density $F_\lambda$.  
In forming model $p$ and $F_\lambda$ spectra, we included a component 
of direct (unpolarized) light of the QSO-like central source 
($F_\nu\propto\nu^{-1}$) that was extinguished by a screen of dust en 
route to the observer, as would occur in an obscuring torus.  We did 
not require an exact fit for either $p$ or $F_\lambda$ in the final 
models because they are both influenced by the contributions of the 
host galaxy and other possible sources of (unpolarized) light, but we 
did constrain them to be reasonable (e.g. $F_\lambda > 0$ and $p \le 
100\%$).

\begin{figure*}
\epsscale{1.5}
\plotone{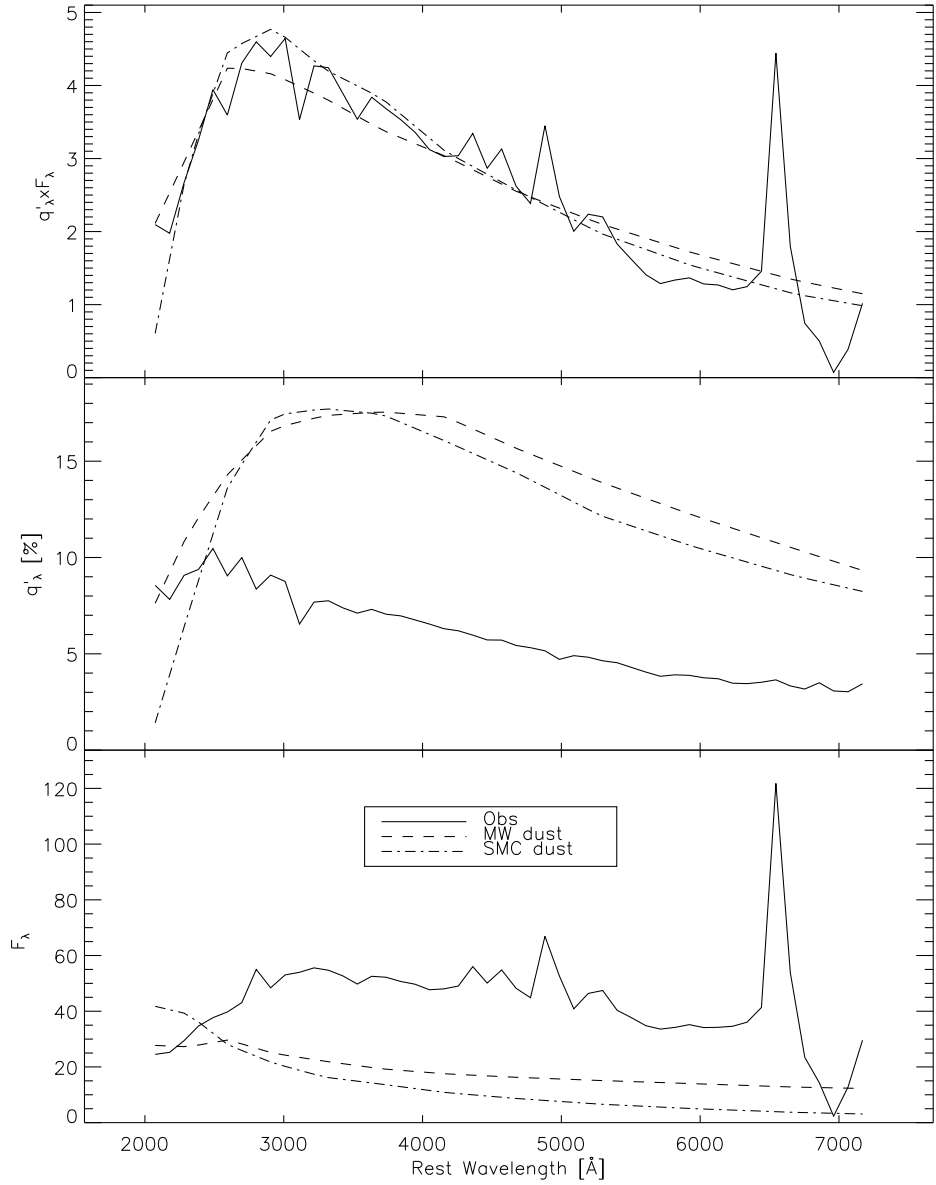}
\figcaption{The best-fitting models for \irasa\ 
assuming Milky-Way (MW) and SMC-type dust in the scattering region.  
For MW scattering dust, the model has a cone opening angle of 
$45\degr$, $i=80\degr$, $\tau_V=1.5$, and the direct light was 
extinguished with a MW-type screen with $E(B-V)=0.60$~mag.  The 
SMC-scattering model assumes an opening angle of $30\degr$, 
$i=90\degr$, $\tau_V=1.5$, and a MW-type screen with 
$E(B-V)=0.30$~mag.}
\end{figure*}

\begin{figure*}
\epsscale{1.5}
\plotone{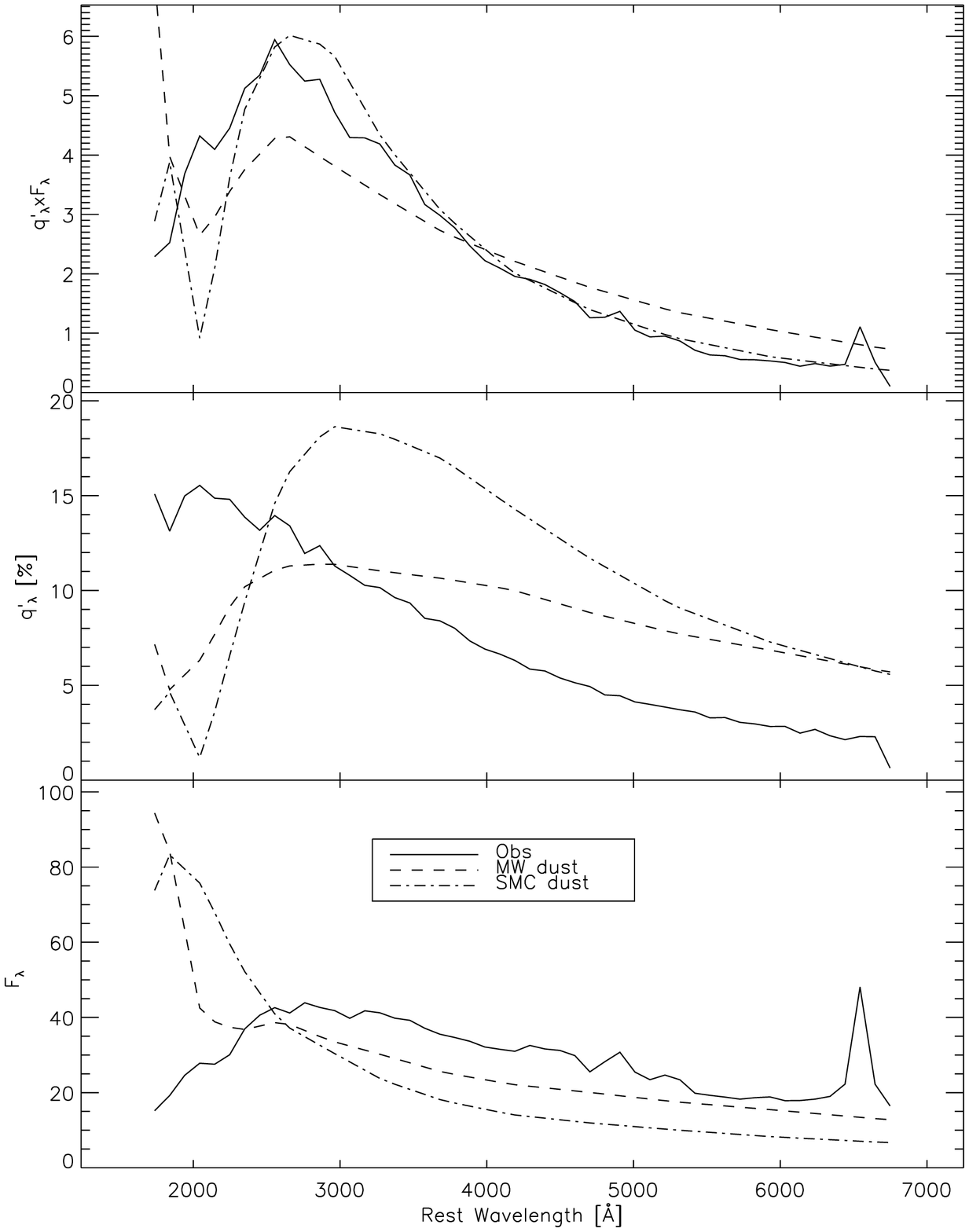}
\figcaption{As in Figure 6 for \irasb.  For MW 
scattering dust, the model has a cone opening angle of $45\degr$, 
$i=90\degr$, $\tau_V=1.0$, and the direct light was extinguished with 
a MW-type screen with $E(B-V)=0.35$~mag.  The SMC-scattering model 
assumes an opening angle of $30\degr$, $i=70\degr$, $\tau_V=0.5$, and 
a MW-type screen with $E(B-V)=0.80$~mag.  }
\end{figure*}

Figures 6 \& 7 show the resulting best-fit models for \irasa\ and\\
\irasb, respectively, with different curves representing MW and 
SMC-type scattering grains in each case.  For \irasa\ we find a 
reasonable fit to the shape of the polarized flux spectrum using 
either MW or SMC-type grains with a radial optical depth $\tau_V=1.5$, 
modest opening angles $\theta_{\rm c}=30\degr-45\degr$, and moderate 
amounts of reddening $E(B-V)=0.3-0.6$.  However, the implied 
inclination is quite large, $i=80\degr-90\degr$.

Only SMC-type grains approximate the observations for \irasb; the best 
model for the polarized flux density spectrum uses $\tau_V=0.5$, 
$\theta_{\rm c}=30\degr$, and $E(B-V)=0.8$.  Once again the derived 
inclination is large, $i=70\degr$.  Yet even for SMC dust, the fit is 
poor below $\sim$3000\AA\ and we were unable to find models which 
simultaneously fit the polarized flux, were above $p_\lambda$, and 
were below $F_\lambda$ at all wavelengths.

For both objects, the models overpredict the degree of polarization by 
about a factor of two, requiring a rather strong diluting component 
over and above the host galaxy starlight.  While this could be 
interpreted as evidence for an additional unpolarized featureless 
continuum as for Seyfert galaxies (the so-called FC2: Tran 1995), the 
similar polarizations of the emission lines and continuum argue 
against a strong FC2 component.  The discrepancy is more likely caused 
by deficiencies in our simple geometry, or in the assumed grain 
properties, or both.  In particular, the very low polarization at 
2000\AA\ in the model for \irasb\ is due to the large number of small 
dust particles required to reproduce the high extinction at short 
wavelengths as deduced for SMC (Gordon \& Clayton 1998).  This effect 
has been illustrated previously in scattering models constructed for 
radio galaxies (Manzini \& di Serego Alighieri 1996).  The presence of 
large numbers of small particles near a QSO nucleus may not be 
realistic since the mid-IR spectra of QSOs do not show the aromatic 
emission bands (commonly called PAH features) that would be emitted by 
small grains heated by UV radiation (e.g., Lutz et al.  1999; 
Rigopoulou et al.  1999; Strum et al.  2000; Tran et al.  2001).  
Planned GTO observations with the {\it Space Infrared Telescope 
Facility (SIRTF)\/} will investigate this possibility for \irasa\ \& 
\irasb\ in particular.

Additional difficulties are posed by the downturn in total spectral 
flux below $\sim$2500\AA, which is seen in both QSOs but does not 
appear in our simple model results.  Possible explanations include an 
intrinsic illuminating spectrum which is heavily reddened before 
reaching the scattering region, reddening of the post-scattered flux, 
or incorrect grain characteristics as mentioned above.  Without 
further observational constraints (e.g., mid-to-far IR spectra), more 
complex models are probably not warranted.

While these simple models are able to reproduce the qualitative 
behavior of the observed polarized flux density spectra, we clearly 
have not converged on a unique solution.  The central QSO light is 
emitted form regions of order the BLR, but the characteristic ``warm'' 
far-IR emission implies a component at a few hundred parsecs.  A 
future enhancement of the code will include adaptive grids to handle a 
wider range of physical scales.  We should then be able to construct a 
self-consitent model that describes both the scattering geometry and 
the thermal emission from the obscuring dust.  This should allow us to 
place stronger constraints on the geometry of the central dusty 
region, especially when combined with detailed mid-to-far infrared 
observations planned with {\it SIRTF}.

\section{DUSTY WARM ABSORBERS \& WIERD DUST}

Given the large reddening of the direct spectrum inferred for \irasa\ 
by Wills et al.  (1992) and assuming a standard Galactic gas-to-dust 
ratio, the soft X-ray emission was originally expected to be highly 
absorbed by cold neutral gas.  Instead, Brandt, Fabian, \& Pounds 
(1996) found that \irasa\ was quite strong in soft X-rays, but showed 
indications of absorption edges due to ``warm'' ionized gas.  This 
prompted the designation of \irasa\ as a ``dusty warm absorber''.  
Warm absorbers have been inferred in many active galaxies by {\it 
ROSAT\/} (e.g., Nandra \& Pounds 1992, Komossa \& Bade 1998) and {\it 
ASCA} (e.g., Kriss et al.  1996, Reynolds 1997).  The X-ray emission 
from these objects is variable on short timescales, implying a direct 
view of the X-ray source, and the spectra typically show features 
often ascribed to absorption edges from \ion{O}{7} and \ion{O}{8}.  
Followup observations of \irasa\ confirmed an absorption feature 
possibly associated with ionized oxygen (Brandt et al.  1997), but the 
feature is not well-defined.  Modeling by Siebert, Komossa, \& 
Brinkmann (1999) failed to explain both the optical and X-ray features 
with a single-zone absorber using a gas-to-dust ratio typical of our 
ISM, so they proposed a model in which the X-rays are absorbed by warm 
gas but the optical extinction is imparted by dust in a physically 
distinct zone.  Complicating the picture, new {\it XMM-Newton} 
observations have revealed strong absorption features from many highly 
ionized species including L-shell excitation of 
\ion{Fe}{17}-\ion{Fe}{20} (Sako et al.  2001), but no clear evidence 
of absorption edges even from \ion{O}{7}.  Nevertheless, the object 
was in an unusually low X-ray flux state, and the reddening inferred 
by Wills et al.  (1992) still cannot be reconciled with the low column 
densities inferred from these new X-ray observations.  Regardless of 
these considerations, all authors point to the importance of UV 
absorption lines for diagnosing the intervening medium, and it is 
significant in this context that we do not detect any absorption 
features in our UV spectrum of \irasa.  We note that the 
ROSAT-detected X-rays observed from \irasb\ (B. Wills, 2001) are 
consistent with the generally weak X-ray emission from BALQSOs (Green 
et al.  1995; Green \& Mathur 1996; Gallagher et al.  1999).

Studies have demonstrated that high polarization is associated with 
the presence of dusty warm absorbers, but the presence of a dusty warm 
absorber does not necessarily imply high polarization (Leighly et al.  
1997; Grupe et al.  1998).  It is also apparent that dusty warm 
absorption and strong optical \ion{Fe}{2} emission are correlated, an 
interesting point considering that all of the objects in Figure 4 are 
extreme optical \ion{Fe}{2} emitters.  The anticorrelation between 
\ion{Fe}{2} and \ion{O}{3} then suggests that polarization and dusty 
warm absorption may be part of ``Eigenvector 1'' (Boroson \& Green 
1992).

Although not formally classified as a warm absorber, Mrk~231 has an 
X-ray spectral shape similar to QSOs (Turner 1999), albeit 
underluminous.  If neutral gas was associated with the high optical 
extinction ($A_{V}=2-7$~ mag: Krabbe et al.  1997 and references 
therein), and Mkn~231 had an intrinsic UV--X-ray spectrum like that of 
unobscured QSOs, the soft X-ray emission would be much less than 
observed.\footnote{These values do not account for any of the 
scattered nuclear light, indicated by the high polarization.  This 
would increase the direct line of sight extinction to be more in line 
with estimates from 10$\mu$m silicate absorption (Rieke 1978; Roche, 
Aitken \& Whitmore 1983) and CO observations (Bryant \& Scoville 
1996).}

The association of high polarization with dusty warm absorption is 
almost surely a result of copious gas and dust near the active nucleus 
which provides both an absorbing screen and a scattering medium, 
though the two need not be co-spatial.  We argued in the last section 
that, even though simple dust scattering models can reproduce the 
general shape of the polarized flux spectra of \irasa\ and \irasb, the 
grain descriptions incorporated into these models might be quite 
unrealistic when applied to the harsh environment near the nucleus of 
a luminous QSO. Lack of neutral absorption in X-rays is a strong 
indication that Galactic dust models may be inapplicable.  Maiolino et 
al.  (2000) and Maiolino, Marconi \& Olivahave (2000) have recently 
revisited the suggestion that (circumnuclear) dust in Seyfert galaxies 
and QSOs may be quite different than in the ISM of our Galaxy.  In 
particular, they find that nearly every object studied has a value of 
$E(B-V)/N_{\rm H}$ smaller than that in our Galaxy by a factor of 3 to 
100, just the opposite of the ``warm absorber'' phenomenon.  
Furthermore, 9.7$\mu$m emission is typically absent in Seyfert 2's 
(e.g., Clavel et al.  2000) and 2175\AA\ absorption is lacking in 
reddened Seyfert 1's.  These authors argue that the circumnuclear dust 
around AGNs is dominated by large grains.  This is also a conclusion 
that we reached from our scattering models.  Thus, while polarization 
by dust scattering is essentially inescapable for the two infrared 
QSOs studied here, the properties of those dust grains are still 
ill-defined.

\section{UNIFIED SCHEMES and BALQSOs}

The discovery of high, wavelength-dependent polarization in \irasa\ 
led Wills et al.  (1992) to a model similar to that for Seyfert 
galaxies (e.g. Antonucci 1993), in which the observed spectrum is the 
combination of QSO light reddened by passage through a dusty torus, 
plus less-reddened, polarized light scattered from within the opening 
of the torus.  This interpretation has been applied successfully to 
many other {\it IRAS\/}-selected AGN (e.g. Hough et al.  1993; 
Antonucci 1993; Young et al.  1996b), including the HIGs (Hines \& 
Wills 1993; Hines et al.  1995, 1999b; Wills \& Hines 1997; Tran, 
Cohen \& Villar-Martin 2000).  At the time of discovery, radio-quiet 
QSOs exhibiting non-time-variable optical polarization like \irasa\ 
occurred exclusively among the few known BALQSOs (Stockman, Moore \& 
Angel 1984).  [More complete surveys confirm that BALQSOs as a class 
are more highly polarized than typical optically-selected QSOs (e.g., 
Hutsemekers, Lamy \& Remy 1998; Schmidt \& Hines 1999).]  Wills et al.  
(1992) predicted that BALQSOs might be common among highly polarized 
{\it IRAS\/}-selected QSOs, since the model for \irasa\ led naturally 
to the idea that normal QSOs could appear as non-BAL or BAL QSOs 
depending on the viewing angle.  Like the Seyfert galaxy paradigm, 
objects viewed at very high inclinations would also be dominated by 
narrow-line emission and host galaxy starlight, while the strong 
featureless continuum, the BELR, \& BAL region (BALR) would be 
obscured from direct view by the dusty torus.

Our subsequent polarization survey of the Low et al.  (1989) sample of 
{\it IRAS\/}-selected QSOs and several HIGs identified significant 
polarization in most of the sample, and confirmed the presence of BALs 
in IRAS 07598+6508 (Wills \& Hines 1995; Schmidt \& Hines 1999) and 
\irasb\ herein (see also Hines \& Wills 1993).  The objects with 
Type-2 optical spectra also revealed hidden broad emission lines as 
expected from the simple unified model (Hines et al.  1995, 1999; 
Young et al.  1996a,b).  Of the original 18 IR-luminous AGN satisfying 
the Low et al.  (1989) ``warm'' far-IR criteria, 15/18 show 
``white-light'' or $V$-band polarization $p>1$\%, and 9/18 have a 
maximum observed polarization (as a function of wavelength) exceeding 
3\% (Hines 1994; Wills \& Hines 1997).  In stark contrast, only 2/114 
PG QSOs have $p>2\%$ (Berriman et al.  1990).  Four of the warm IR 
QSOs are now confirmed BALQSOs, and all of these are highly polarized 
($p\ge2\%$).  It may then be interesting that despite its high 
polarization and X-ray absorption lines, that we have not found BALs 
in \irasa.

While \irasa\ and \irasb\ are similar in many ways, key differences 
can also be noted.  \irasa\ is: 1) a (weak) [\ion{O}{3}] $\lambda5007$ 
emitter; 2) a significantly weaker \ion{Fe}{2} emitter as evidenced by 
the H$\beta$/\ion{Fe}{2}(4500) ratio; and 3) a relatively strong X-ray 
emitter.  It is tempting to speculate that \irasa\ is simply viewed at 
a smaller inclination so our direct line of sight does not pass 
through the BAL region or X-ray absorbing material.  If the 
\ion{Fe}{2}-emitting region is located outside both the BELR and BALR, 
then the H$\beta$/\ion{Fe}{2}(4500) ratio might be smaller at lower 
inclinations (see, e.g., Keel et al.  1994).  A similar effect could 
be invoked for the [\ion{O}{3}] emission if a significant fraction of 
the [\ion{O}{3}]-emitting gas is obscured at high inclinations as 
appears to be the case for some radio galaxies (Hes, Barthel \& 
Fosbury 1993; di Serego Alighieri et al.  1997) and at least one HIG 
(Tran, Cohen \& Villar-Martin 2000).  On the other hand, there are 
indications that [\ion{O}{3}] emission is isotropic in radio-quiet 
objects (Kuraszkiewicz et al.  2000).

An alternative scenario to simple orientation dependence may apply.  
Sanders et al.  (1988) suggested that QSOs ``evolve'' from 
Ultraluminous Infrared Galaxies dominated by starbursts into optically 
bright, UV excess (-selected) QSOs.  If AGNs are fueled by galaxy 
collisions, then an initial collision between gas-rich spiral galaxies 
could fuel a nascent AGN. As the dust settles, the QSO nucleus would 
evaporate away its cocoon of dusty gas, where material would be 
ablated and accelerated away from the nucleus to form the BALR. The 
dusty cocoon might acquire a toroidal geometry and emit strongly in 
the mid-IR. The dust in the cocoon and surrounding region would supply 
the scattering material we infer from the high polarization, and 
compete for UV photons that would otherwise ionize the surrounding 
gas.  This suggests that the lack of [\ion{O}{3}] emission in BALQSOs 
and the strength of \ion{Fe}{2} might primarily be manifestations of 
the dust covering factor (e.g., Boroson \& Meyers 1992) as opposed to 
simple orientation alone.

All of the objects in Figure 4 have very strong \ion{Fe}{2} emission 
and extremely weak [\ion{O}{3}] emission.  In addition, the two 
objects which do show slight evidence for the optical [\ion{O}{3}] 
lines do not show obvious signs of BALs in the UV or optical (\irasa\ 
\& Mrk~486).  Finally, all three BALQSOs are members of the rare class 
of low-ionization BALQSOs.  Boroson \& Meyers (1992) have suggested 
that the low-ionization BALQSOs have a higher covering factor of 
absorbing material than QSOs that show only high-ionization BALs.  The 
tendency for low-ionization BALQSOs to appear reddened (e.g., 
Sprayberry \& Foltz 1992), more highly polarized (Hutsemekers, Lamy \& 
Remy 1998; Schmidt \& Hines 1999), and perhaps show strong evidence 
for dust scattering all support this idea.

Unfortunately, the limited sensitivity of {\it IRAS\/} enabled the 
identification of only a handful of extremely luminous infrared AGN, 
so separating evolution from orientation is quite difficult.  
Furthermore, very few mid-to-far IR observations are available since 
most BALQSOs are at $z\ge1$.  New observations, using especially the 
Multiband Imaging Photometer for {\it SIRTF\/} (MIPS) will enable us 
to look for differences in the mid-to-far IR spectral energy 
distributions (SEDs) that may indicate changes in covering factor: the 
mid-to-far IR SEDs should be less affected by orientation, but should 
be sensitive to dust cover (but see Pier \& Krolik 1992; Efstathiou \& 
Rowan-Robinson 1995).  Spectral features observed with the Infrared 
Spectrograph aboard {\it SIRTF\/} may also allow constraints on the 
intrinsic (ionizing) continuum spectrum.

In the meantime, what can we do to investigate these issues further?  
Optical surveys for QSOs are biased against highly obscured objects, 
and thus do not accurately represent the parent population of QSOs 
(e.g., Wills \& Hines 1997).  The results discussed earlier for the 
{\it IRAS\/}-selected QSOs/HIGs emphasized this and showed that many 
QSOs may be revealed only by their strong infrared emission and in 
light polarized by scattering.  Importantly, all of the warm objects 
which did not show the QSO directly have strong forbidden narrow-line 
emission from highly ionized species, indicating that at least some 
lines of sight from the QSO are sufficiently unobscured to allow 
ionizing photons from the nucleus to reach the narrow-line gas.  
Indeed, many new luminous mis-directed AGN are being identified in 
color-selected surveys when strong [\ion{O}{3}] alters the broad band 
colors (e.g. Djorgovsky et al.  2000).

Cutri et al.  (2001, in prep) have used the accurate ($JHK$) 
photometry and astrometry of the The Two Micron All Sky Survey (2MASS: 
Beichman et al.  1998) to locate previously unidentified, red QSOs at 
$z\le0.7$.  These 2MASS QSOs tend to be very highly polarized (Smith 
et al.  2000), with both polarization and far-IR properties 
intermediate between the BALQSOs and {\it IRAS\/}-QSOs/HIGs.  As for 
the {\it IRAS\/}-QSOs, these objects may simply be more highly 
inclined PG-type QSOs, or they may have a higher dust cover.  In 
either case, we might then expect a large fraction of low-ionization 
BALQSOs in the 2MASS sample.  Both UV and IR-spectroscopy will be 
essential in gaining an understanding of this new, important sample of 
objects.

The sensitivity of {\it IRAS} limited far-IR color selection to 
objects at $z < 1$.  {\it SIRTF} should provide about three orders of 
magnitude increase in sensitivity, so should be able to indentify warm 
AGN to redhifts of 10 if they exists (Low \& Hines 1998; Hines \& Low 
1999).  Since obscured AGN equal or outnumber UV-selected QSOs in the 
local universe, we may expect that objects like the 2MASS QSOs and 
{\it IRAS\/}-QSOs/HIGs will dominate the early universe.  Deep surveys 
planned with {\it SIRTF} should provide many examples and allow us to 
probe both galaxy and QSO formation at the earliest epochs.

Finally, even more obscured objects may be hard to identify by 
color-selection alone, even in the far-IR. For these objects, hard 
X-ray emission may be the key discriminator (e.g. Almaini, Lawrence \& 
Boyle 1999; Barger et al.  2001).  Missions such as {\it Chandra} and 
{\it XMM} have already discovered several previously unidentified, 
optically obscured AGNs (e.g. Fiore et al.  2000).

\section{SUMMARY \& CONCLUSIONS}

Ground-based optical and {\it HST}/FOS UV spectropolarimetry of 
\irasa\ and \irasb\ reveals polarized flux density spectra that 
strongly indicate scattering by dust grains as the polarizing 
mechanism.  We have shown that dust scattering in a geometry like that 
inferred for simple Unified Schemes is able to qualitatively reproduce 
the polarized flux density spectra for these and other {\it 
IRAS\/}-selected AGN. However, we also suggest that the grain 
distribution may be skewed toward larger particles than are typically 
found in the ISM of our Galaxy or in the Small Magellanic Cloud.  
Support for this inference is found in both X-ray and mid-IR results 
for AGN.

While these results may bid farewell to the most simplistic 
interpretations of polarization spectra, they offer hope that 
polarized flux distributions might reveal the detailed properties of 
material illuminated by the central engine.  The primary uncertainty 
is the exact form of the intrinsic spectrum that illuminates the 
scattering material.  If the shape of this intrinsic spectrum can be 
determined (e.g., via mid-to-far IR emission lines as suggested by 
Spinoglio \& Malkan 1992; Voit 1992), then strong constraints can be 
placed on the properties of the scattering material.  The {\it 
SIRTF\/} mission will provide the first opportunity to obtain the 
spectra necessary to pursue such an investigation of these objects.

Our UV spectropolarimetry also confirms that \irasb\ is a 
low-ionization BALQSO, but perhaps surprisingly, we find no indication 
of absorption lines in \irasa.  The difference may be orientation or 
covering factor of both the dust and the BAL material.  Again, 
spectroscopic observations with {\it SIRTF\/} should assist us in 
untangling these possibilities.  The fact that polarimetry at short 
wavelengths is necessary to distinguish dust scattering from electron 
scattering (as demonstrated herein) cries out for UV 
spectropolarimetric capability.  Unfortunately, the only platform on 
the horizon -- the Advanced Camera for Surveys (ACS) for \HST\ -- will 
not provide sufficient spectral resolution with its Grisms for 
detailed analysis of the polarization properties of the UV emission 
and absorption lines of AGN. A space-based UV capability is needed.

\acknowledgements

It is a pleasure to thank R. Lucas and A. Koratkar for assistance at 
STScI with FOS observations.  Thanks also to R. Goodrich and D. Doss 
for assistance with the McDonald Spectropolarimeter.  We have 
benefited from illuminating discussions with F. Low and M. Brotherton.  
We especially acknowledge Vino Vat's (Karl Lambrecht Corporation) 
contribution to astronomical spectropolarimetry, especially at 
McDonald Observatory.  We thank the referee, R.R.J. Antonucci for his 
critical reading of the manuscript and for his suggestions which have 
improved the paper.  Support for this work was provided by NASA 
through grant number GO-5829-01-94A from the Space Telescope Science 
Institute, which is operated by Association of Universities for 
Research in Astronomy, Incorporated, under NASA contract NAS5-26555.  
DCH acknowledges additional support from NASA grant NAG5-53359 to US 
{\it ISO} observers.  DCH \& KG acknowledge support from the MIPS 
Project (MIPS Science Development is supported by the National 
Aeronautics and Space Administration (NASA) and the Jet Propulsion 
Laboratory (JPL) under Contract 960785 to the University of Arizona).

\clearpage

\begin{deluxetable}{llclrccr}
\scriptsize
\tablenum{1}
\tablecaption{Observation Log}
\tablewidth{0pc}
\tablehead{\colhead{Object}
& \colhead{Date}
& \colhead{Telescope}
& \colhead{Inst.}
& \colhead{Slit}
& \colhead{Time}
& \colhead{Range} &
\colhead{$\Delta\lambda$} \\
\colhead{(IRAS)}
& \colhead{(UT)}
& \colhead{}
& \colhead{}
& \colhead{(sec)}
& \colhead{\arcsec}
& \colhead{(\AA)} &
\colhead{(\AA)}}
\startdata
13349+2438 & 1996 May 08    & \HST\   & FOS/G190H & 1.0 &1670  & 1573$-$2329 & 1.5 \\
        & 1996 May 08        & \HST\   & FOS/G270H/Pol & 1.0 & 6360 & 2214$-$3302 & 2.1 \\
        & 1991 Mar 17        & MO 2.7m & LCS/SPOL  & 2-3 & 7200  & 5197$-$7922 & 6.9 \\
        & 1992 Mar 05        & MO 2.7m & LCS/SPOL  & 2-3 & 7200  & 3063$-$5775    & 6.9 \\
        & 1995 Mar 17        & SO 2.3m & CCD SPOL  & 2-3 & 4800  & 3990$-$8000    & 12 \\
        &                    &            &           &       &                 & \\
14026+4341 & 1994 Jul 17\tablenotemark{a} & \HST\   1.0 & FOS/G190H & 978  & 1573$-$2329 & 1.5 \\
        & 1995 Sep 11      & \HST    & FOS/G270H/Pol & 1.0 & 12,590 & 2214$-$3302 & 2.1 \\
        & 1992 Mar 17,18   & MO 2.7m & LCS/SPOL  & 2-3 & 9600  & 5197$-$7922  & 6.9 \\
        & 1992 Mar 05,06   & MO 2.7m & LCS/SPOL  & 2-3 & 16,500 & 3063$-$5775 & 6.9 \\
        & 1996 Mar 19,22   & SO 2.3m & CCD SPOL  & 2-3 & 16,800 & 4760$-$9000 & 12 \\
\enddata
\tablenotetext{a}{Data from STScI Archive.  We gratefully
acknowledge the original proposers and observers, Turnshek et al.
(1996); GO 5456 (PI: D. Turnshek).}

\end{deluxetable}

\begin{deluxetable}{lcccccc}
\tablenum{2}
\tablewidth{0pt}
\tablecaption{Polarization of Spectral Features\tablenotemark{a}}
\tablehead{
\colhead{Object}
& \colhead{$\Delta\lambda$}
& \colhead{$p$}
& \colhead{$\sigma_{p}$}
& \colhead{$\Theta$}
& \colhead{$\sigma_{\theta}$}
& \colhead{Feature} \\
\colhead{}
& \colhead{(\AA)}
& \colhead{($\%$)}
& \colhead{($\%$)}
& \colhead{(\deg)}
& \colhead{(\deg)}
& \colhead{}}
\startdata
IRAS 13349+2438& 2008$-$2708 & 9.44&         0.24&  123.7&           0.8&Cont.           \\
              & 2708$-$3611 & 7.93&         0.17&  125.0&           0.6&Cont.           \\
              & 3613$-$4223 & 6.65&         0.04&  123.0&           0.2&Cont.           \\
              & 4259$-$4385 & 7.43&         0.70&  124.9&           2.7&H$\gamma$       \\
              & 4696$-$4800 & 5.31&         0.05&  122.6&           0.3&Cont.           \\
              & 4790$-$4920 & 4.35&         0.20&  118.6&           1.1&H$\beta$        \\
              & 4974$-$5032 & 2.16&         0.80&  134.2&          10.6&[\ion{O}{3}] $\lambda$5007\\
              & 5046$-$5129 & 4.89&         0.05&  123.1&           0.3&Cont.           \\
              & 5461$-$6429 & 3.78&         0.02&  124.8&           0.2&Cont.           \\
              & 6436$-$6674 & 3.67&         0.05&  119.1&           0.4&H$\alpha$       \\
              & 6692$-$7223 & 3.24&         0.04&  123.7&           0.3&Cont.           \\
              &                 &     &             &       &              &                \\
IRAS 14026+4341& 1687$-$2270 &14.84&         0.22&   30.5&           0.5&Cont.           \\
              & 2270$-$2729 &13.67&         0.26&   30.1&           0.6&Cont.           \\
              & 2781$-$2832 & 7.16&         2.90&   14.4&          11.6&\ion{Mg}{2} $\lambda$2800\\
              & 2872$-$3775 & 8.63&         0.04&   29.3&           0.2&Cont.           \\
              & 3781$-$4280 & 6.71&         0.02&   30.0&           0.1&Cont.           \\
              & 4316$-$4365 & 5.89&         1.00&   25.3&           4.9&H$\gamma$       \\
              & 4679$-$4794 & 4.73&         0.04&   31.1&           0.3&Cont.           \\
              & 4815$-$4903 & 3.45&         0.45&   24.2&           3.7&H$\beta$        \\
              & 5396$-$6381 & 3.11&         0.02&   30.7&           0.2&Cont.           \\
              & 6417$-$6662 & 1.86&         0.30&   33.7&           4.6&H$\alpha$       \\
              & 6662$-$6804 & 2.66&         0.50&   25.8&           5.3&Cont.           \\
\enddata

\tablenotetext{a}{All measurements in the rest frame. Line measurements made 
after subtraction of local continuum.}

\end{deluxetable}

\begin{deluxetable}{llccccc}
\tablenum{3}
\tablewidth{0pt}
\tablecaption{Emission and Absorption Line Properties\tablenotemark{a}}
\tablehead{
\colhead{Object}
& \colhead{Feature}
& \colhead{$\lambda$}
& \colhead{FWHM}
& \colhead{$\Delta v$}
& \colhead{EW}
& \colhead{Observed Flux} \\
\colhead{({IRAS})}
& \colhead{}
& \colhead{(\AA)}
& \colhead{(\AA)}
& \colhead{(km~s$^{-1}$)}
& \colhead{(\AA)}
& \colhead{($\times10^{-13}$ erg~cm$^{-2}$~s$^{-1}$)}}
\startdata
P13349+2438 &   \ion{C}{4}    &   1549    &   25  &   4839  &  25  &   4.40    \\
                &   \ion{C}{3}]\tablenotemark{b}  &   1903    &   28  &   4411  &  17  &   3.67    \\
                &   \ion{Mg}{2}$_{\rm em,b}$\tablenotemark{c}
                            &   2783    &   106 &   11419 &  12  &   5.13    \\
                &   \ion{Mg}{2}$_{\rm em,n}$
                            &   2796    &   24  &   2573  &  14  &   5.13    \\
                &   \ion{Mg}{2}$_{\rm em,b+n}$
                            &  \ldots   & \ldots & \ldots  &  26  &   10.26   \\
                &   H$\delta$+\ion{N}{3}+\ion{He}{2}
                            &   4101    &   49  &   3582  &  6   &   2.20    \\
                &   H$\gamma$+[\ion{O}{3}]
                            &   4347    &   45  &   3104  &  19  &   6.60    \\
                &   \ion{Fe}{2}   & 4450$-$4700 & \ldots &  \ldots &  63  &   25.66   \\
                &   H$\beta$&   4861    &   49  &   3021  &  53  &   19.80   \\
                &   \ion{Fe}{2}   & 4750$-$5100 &  \ldots &   \ldots  &  43  &   16.13   \\
                &   [\ion{O}{3}] &   4959    &   20  &   1209  &  2   &   0.73    \\
                &   [\ion{O}{3}] &   5007    &   23  &   1377  &  8   &   2.93    \\
                &   \ion{Fe}{2}   & 5100$-$5600 & \ldots&  \ldots &  81  &   27.13   \\
                &   H$\alpha$+[\ion{N}{2}]
                            &   6563  &   55  &   2512  &  406  &   115.8  \\
                &           &           &       &         &      &          \\
P14026+4341 &   \ion{C}{4}$_{\rm abs, detached}$
                            &   1432    &   54  &   11305 &  39  &   3.67    \\
                &   \ion{C}{4}$_{\rm abs}$
                            &   1519    &   8   &   1579  &  5   &   0.73    \\
                &   \ion{C}{4}$_{\rm em}$ (?)
                            &   1542    &   6   &   1167  &  2   &   0.22    \\
                &   \ion{Al}{3}$_{\rm abs}$
                            & 1855$+$1863 &   18  &   2931  &  8   &   2.20    \\
                &   \ion{C}{3}]+\ion{Fe}{3}\tablenotemark{b}
                            &   1905    &   56  &   8813  &  10  &   2.93    \\
                &    (?)$_{\rm em}$
                            &   2075    &   47  &   6791  &  9   &   2.93    \\
                &   Galactic \ion{Mg}{2}$_{\rm abs}$
                            &   2116\tablenotemark{d}
                                        & \ldots &  \ldots&\ldots& \ldots    \\
                &    (?)$_{\rm em}$
                            &   2425    &   25  &   3091  &  3   &   1.47    \\
                &   \ion{He}{2}+[\ion{Mg}{7}]
                            &   2512    &   27  &   3222  &  2   &   1.47    \\
                &    (?)$_{\rm abs}$
                            &   2658    &   33  &   3722  &  3   &   1.47    \\
                &    \ion{Mg}{2}$_{\rm em}$
                            &   2799    &   36\tablenotemark{e}
                                                &   3856  &18\tablenotemark{e}
                                                                 &   9.53    \\
                &   \ion{Mg}{2}$_{\rm abs}$
                            &   2770    &   27    &   2922  & 7    &   3.67    \\
                &   H$\delta$+\ion{N}{3}+\ion{He}{2}
                            &   4107    &   40  &   2920  &  2   &   0.73    \\
                &   H$\gamma$+[\ion{O}{3}]
                            &   4346    &   34  &   2345  &  6   &   2.20    \\
                &   \ion{Fe}{2}    & 4450$-$4700 &\ldots & \ldots  & 48   &   15.40   \\
                &  H$\beta$ &   4861    &   48  &   2960  &  32  &   9.53    \\
                &   \ion{Fe}{2}    & 4750$-$5100 & \ldots&   \ldots&  35  &   10.26   \\
                &   \ion{Mg}{2}   & 5100$-$5600 & \ldots&  \ldots &  62  &   15.40   \\
                &   H$\alpha$+[\ion{N}{2}]
                            &   6566    &   59  &   2694  &  318\tablenotemark{c} &   63.05   \\
\enddata

\tablenotetext{a}{All measurements in the rest frame.  For both
objects, \ion{Fe}{2} was measured using a broadened \ion{Fe}{2}
template (I Zw1) from Boroson \& Green (1992).}

\tablenotetext{b}{Possible contamination from poorly modeled broad \ion{Si}{3}].}

\tablenotetext{c}{Possible contamination from poorly modeled broad \ion{Fe}{2}.}

\tablenotetext{d}{Appears in the spectrum at 2116\AA.}

\tablenotetext{e}{Intrinsic emission line assumed to be symmetric.}

\end{deluxetable}

\end{document}